\documentclass[aps,twocolumn,nofootinbib,floatfix,superscriptaddress]{revtex4}
\usepackage{amsfonts}
\usepackage{amssymb}
\usepackage{amsmath}
\usepackage{graphics}
\usepackage{graphicx}
\usepackage{soul}
\usepackage{natbib}
\usepackage{bm}
\usepackage{mathtools}
\usepackage{epstopdf}
\usepackage{color}
\usepackage{lipsum}
\usepackage{balance}
\usepackage[colorlinks=true,citecolor=black,linkcolor=black,urlcolor=black,filecolor=black]{hyperref}

\usepackage{acronym}
\newacro{PDF}{probability distribution function}
\newacroplural{PDF}[PDFs]{probability distribution functions}
\newcommand{\PDF}{\ac{PDF}}
\newcommand{\PDFs}{\acp{PDF}}
\newacro{DF}{distribution function}
\newacroplural{DF}[DFs]{distribution functions}
\newcommand{\DF}{\ac{DF}}
\newcommand{\DFs}{\acp{DF}}
\newacro{HMF}{Hamiltonian Mean Field}
\newcommand{\HMF}{\ac{HMF}}
\newacro{GPU}{graphics processing unit}
\newacroplural{GPU}[GPUs]{graphics processing units}
\newcommand{\GPU}{\ac{GPU}}
\newcommand{\GPUs}{\acp{GPU}}

\newcommand{\rd}{\mathrm{d}}
\newcommand{\re}{\mathrm{e}}
\newcommand{\ri}{\mathrm{i}}
\newcommand{\mP}{\mathcal{P}}
\newcommand{\deltaD}{\delta_{\mathrm{D}}}
\newcommand{\veps}{\varepsilon}
\newcommand{\ot}{\overline{t}}
\newcommand{\td}{t_{\mathrm{d}}}
\newcommand{\tr}{t_{\mathrm{r}}}
\newcommand{\Mtot}{M_{\mathrm{tot}}}
\newcommand{\oF}{\overline{F}}
\newcommand{\Umax}{U_{\mathrm{max}}}
\newcommand{\oU}{\overline{U}}
\newcommand{\omU}{\overline{\mathcal{U}}}
\newcommand{\mU}{\mathcal{U}}
\newcommand{\bk}{\mathbf{k}}
\newcommand{\bvel}{\mathbf{v}}
\newcommand{\FB}{F_{\mathrm{B}}}
\newcommand{\kp}{k^{\prime}}
\newcommand{\np}{n^{\prime}}

\newcommand{\mR}{\mathcal{R}}
\newcommand{\mF}{\mathcal{F}}
\newcommand{\mC}{\mathcal{C}}
\newcommand{\omC}{\overline{\mathcal{C}}}
\newcommand{\obk}{\overline{\mathbf{k}}}
\newcommand{\kpsq}{k^{\prime 2}}
\newcommand{\mO}{\mathcal{O}}
\newcommand{\wbM}{\widehat{\mathbf{M}}}
\newcommand{\bveps}{\bm{\veps}}
\newcommand{\bI}{\mathbf{I}}
\newcommand{\oom}{\overline{\omega}}
\newcommand{\Qc}{Q_{\mathrm{c}}}

\newcommand{\Nthreads}{N_{\mathrm{threads}}}
\newcommand{\Nblocks}{N_{\mathrm{blocks}}}
\newcommand{\Nruns}{N_{\mathrm{runs}}}
\newcommand{\omF}{\overline{\mathcal{F}}}
\newcommand{\bu}{\mathbf{u}}
\newcommand{\kmax}{k_{\mathrm{max}}}
\newcommand{\bw}{\mathbf{w}}

\begin{document}

\title{Kinetic theory of one-dimensional homogeneous long-range interacting systems\\
with an arbitrary potential of interaction}

\author{Jean-Baptiste Fouvry}
\affiliation{CNRS and Sorbonne Universit\'e, UMR 7095, Institut d'Astrophysique de Paris,\\98 bis Boulevard Arago, F-75014 Paris, France}
\author{Pierre-Henri Chavanis}
\affiliation{Laboratoire de Physique Th\'eorique, Universit\'e de Toulouse, CNRS, UPS, France}
\author{Christophe Pichon}
\affiliation{CNRS and Sorbonne Universit\'e, UMR 7095, Institut d'Astrophysique de Paris,\\98 bis Boulevard Arago, F-75014 Paris, France}
\affiliation{Korea Institute for Advanced Study, 85 Hoegiro, Dongdaemun-gu, 02455 Seoul, Republic of Korea}

\begin{abstract}
Finite-$N$ effects unavoidably drive the long-term evolution
of long-range interacting $N$-body systems.
The Balescu-Lenard kinetic equation generically describes
this process sourced by ${1/N}$ effects
but this kinetic operator exactly vanishes by symmetry for one-dimensional
homogeneous systems: such systems undergo a kinetic blocking
and cannot relax as a whole at this order in ${1/N}$.
It is therefore only through the much weaker ${1/N^{2}}$ effects,
sourced by three-body correlations,
that these systems can relax, leading to a much slower evolution.
In the limit where collective effects can be neglected,
but for an arbitrary pairwise interaction potential,
we derive a closed and explicit kinetic equation
describing this very long-term evolution.
We show how this kinetic equation satisfies an $H$-theorem
while conserving particle number and energy,
ensuring the unavoidable relaxation of the system towards
the Boltzmann equilibrium distribution.
Provided that the interaction is long-range,
we also show how this equation cannot suffer from further kinetic blocking,
i.e.\@, the ${1/N^{2}}$ dynamics is always effective.
Finally, we illustrate how this equation quantitatively matches
measurements from direct $N$-body simulations.
\end{abstract}
\maketitle

\section{Introduction}
\label{sec:Introduction}

The statistical mechanics and kinetic theory of systems with long-range interactions is a
topic of great interest~\cite{campabook} because of its unusual properties
(ensembles inequivalence, negative specific heats, non-Boltzmannian quasistationary
states, instabilities, phase transitions...) and its applications in various domains
of physics such as plasma physics~\cite{nicholson}, astrophysics~\cite{BinneyTremaine2008},
or two-dimensional hydrodynamics~\cite{houches,bv}.

Closed systems with long-range interactions generically experience two successive types of relaxations.
There is first a fast collisionless relaxation
driven by the mean field towards a non-Boltzmannian quasistationary state.
This corresponds to the process of violent relaxation described by Lynden-Bell~\cite{lb}
for collisionless stellar systems governed  by the Vlasov-Poisson equations
(see, e.g.\@,~\cite{csr}). This phase takes place within a few dynamical times (independent of the number
of particles) and ends when the system has reached a virialized state,
i.e.\@, a stable steady state of the Vlasov equation.
Then, a slow collisional relaxation towards
the Boltzmann distribution of statistical equilibrium takes place. It is
driven by discreteness effects (granularities) due to finite values of $N$,
the total number of particles.
The relaxation time
expressed in units of the dynamical time diverges with the number of
particles $N$.\footnote{For stellar systems $N$
represents the number of stars in the system (or the number of stars in
the Jeans sphere $(n\lambda_{\mathrm{J}}^3)$); in plasma physics $N$ represents the number
of ions in the Debye sphere $(n\lambda_{\mathrm{D}}^{3})$.} In this sense, the lifetime of the
quasistationary state becomes infinite when $N\rightarrow +\infty$. Nevertheless, for large
but finite values of $N$, the system evolves secularly, passing adiabatically by a succession
of quasistationary states.

The derivation of kinetic equations describing the secular evolution of systems with
long-range interactions has a rich history (see, e.g.\@, the introduction of~\cite{aa,epjp2}
for a short account).  Landau~\cite{landau} first derived a kinetic equation for Coulombian neutral plasmas
by expanding  the Boltzmann~\cite{boltzmann} equation in terms of a small deflection parameter,
namely the velocity deviation experienced by a particle during a ``collision''.
An equivalent kinetic equation  was obtained independently by Chandrasekhar~\cite{chandra}
(and generalized by Rosenbluth \textit{et al.}~\cite{rosenbluth}) for stellar systems.
Chandrasekhar started  from the Fokker-Planck equation and calculated the diffusion and friction
coefficients using an impulse approximation. However, the approaches of Landau and Chandrasekhar
have a phenomenological character and ignore collective effects and spatial inhomogeneity.
This leads to difficulties such as the logarithmic
divergence of the collision term at large impact parameters.

Systematic and rigorous approaches directly starting from the $N$-body dynamics
(or from the Liouville equation) were  developed by Bogoliubov~\cite{bogoliubov}
using a hierarchy  of equations for the reduced distribution functions (nowadays
called the BBGKY hierarchy)  and by Prigogine and Balescu~\cite{pbdt} using diagrammatic techniques.
These hierarchies of equations may be closed by considering an expansion of the equations
in powers of the small coupling parameter $1/N$ (with ${ N \gg 1 }$) which measures the
strength of the correlation functions. Initially, only two-body correlation functions,
which are of order $1/N$, were taken into account. This corresponds to the weak coupling
approximation of plasma physics. These methods  led to the Balescu-Lenard
equation~\cite{balescu,lenard} which takes into account collective effects (dynamical Debye shielding) thereby removing the
logarithmic divergence that occurs in the Landau equation at large scales.
This kinetic equation describes the effect of two-body encounters and is essentially exact
at order $1/N$. It can also be derived from a quasilinear theory
based on the Klimontovich equation for the discrete distribution function~\cite{klimontovich}.
The original Balescu-Lenard equation (applying to neutral plasmas) is valid
for spatially homogeneous systems but it has recently been generalized to
inhomogeneous systems by using angle-action variables~\cite{heyvaerts,physicaA} with
specific applications to self-gravitating systems~\cite{fpmc,fpcm,hfbp,bf}
and to the  magnetized phase of the \HMF\ model~\cite{bm}. More
generally,  the Balescu-Lenard kinetic equation is valid for any system with long-range
interactions in arbitrary dimension of space~\cite{epjp1}.
For usual three-dimensional (3D)
systems, this kinetic equation conserves the particle number and the energy,
and satisfies
an $H$-theorem for the Boltzmann entropy. As a result, it relaxes towards the Boltzmann
distribution which is the maximum entropy state  (most  probable state) at fixed particle
number and energy. Since the Balescu-Lenard equation is valid at order $1/N$ it describes
the relaxation of the system on a timescale of order $N \td$,
with $\td$ the dynamical time.
Actually, for Coulombian plasmas and stellar systems, there is a logarithmic correction due to strong
collisions at small impact parameters,
so that the relaxation
time scales as ${ (N/ \ln N) \td }$.

Apart from specificities inherent to systems with long-range
interactions (the process of violent relaxation, the existence of transient non-Boltzmannian
quasistationary states, the very long relaxation time, the need to account for
spatially  inhomogeneous distributions, and the importance of collective effects)
the results of the kinetic theory at order $1/N$ are consistent with the
original Boltzmann picture of relaxation in a dilute gas. In a sense, the Balescu-Lenard
equation (and more specifically the homogeneous Landau equation) is a descendent
of the Boltzmann equation: the collision term is the product of two distribution functions
${ F(1) F(2) }$ characteristic of any two-body collision term and the derivation of
the conservation laws and of the $H$-theorem
is essentially the same as that given by Boltzmann.\footnote{The Balescu-Lenard
equation exhibits a new type of nonlinearity which is directly related to the
collective nature of the interaction, but this does not affect the derivation
of the conservation laws and of the $H$-theorem.}

However, for 1D homogeneous systems, the Balescu-Lenard
collision term vanishes identically.\footnote{This is also the case for the Boltzmann
and Landau collision terms. By contrast, for one dimensional inhomogeneous systems,
the Balescu-Lenard and  Landau collision terms written with angle-action variables are non-zero.}
As a result, the dynamics sourced by two-body correlations is frozen at order ${ 1/N }$,
so that there is no evolution of the  overall distribution on a timescale ${ N \td }$.
This is a situation of kinetic blocking. The system is therefore expected
to evolve dynamically under the effect  of nontrivial three-body (or higher) correlations
implying that the relaxation time should scale as ${ N^2 \td }$ (or be even larger).\footnote{We
shall prove in this paper that the relaxation time is never larger than ${ N^2 \td }$,
for long-range interactions.}
The peculiarity of 1D homogeneous systems was first noticed by Eldrige and Feix~\cite{ef}
in the context of 1D plasmas~\cite{dawson1}. They showed that the
Balescu-Lenard collision term vanishes and
conjectured the existence of a non-zero ${ 1/N^2 }$ collision term. The corresponding ${ N^2 \td }$
scaling of the relaxation time was confirmed by Dawson~\cite{dawson2} from direct $N$-body
simulations. Later, Rouet and Feix~\cite{rf} illustrated the striking difference that exists
between the relaxation of the system as a whole (overall distribution) which takes place
on a timescale ${ N^2 \td }$ and the relaxation of test (or labelled) particles which takes
place on a timescale ${ N \td }$. The stochastic evolution of the test particles is governed
by a Fokker-Planck equation which can be obtained from the Balescu-Lenard equation by
making a bath approximation, i.e.\@, by fixing the distribution of the field particles. This procedure
transforms an integro-differential equation into a differential equation.
Since in 1D the test particles acquire the distribution of the field particles
(bath) whatever its distribution function (while this is true only for the Boltzmann
distribution in 3D) this explains why a 1D homogeneous system does not evolve on a timescale
${ N \td }$.

Similar results were found later for axisymmetric distributions of 2D point
vortices when the profile of angular velocity ${ \Omega(r,t) }$ is monotonic~\cite{dubin1,pv1,dubin2,pv2,pv3,pv4} and for 1D systems with long-range
interactions such as the \HMF\ model~\cite{bd,cvb} and classical spin systems
with anisotropic interaction (or equivalently long-range interacting particles
moving on a sphere)~\cite{gm1,bg,lr,fbc}.
In the context of the \HMF\ model, it was first believed that the relaxation time was
anomalous, scaling with the number of particles as ${ N^{1.7} \td }$~\cite{yamaguchi}.
However, it was later demonstrated~\cite{rocha1,rocha2,fbcN2} that this anomalous exponent was
due to small size effects and that the correct scaling is indeed ${ N^{2} \td }$ in agreement
with kinetic theory~\cite{epjp1}.\footnote{Similar results were obtained for spin
systems in \cite{lr,fbc}.} The collisional relaxation of the \HMF\ model was studied by~\cite{ccgm}
who found that, for certain initial conditions, the distribution
function ${ F (v,t) }$ can be fitted by polytropes with a time-dependent index. When the
polytropic index reaches a critical  value, the distribution function
becomes dynamically unstable (with respect to the Vlasov equation) and a dynamical phase
transition from a homogeneous phase to an inhomogeneous phase takes place. These authors
stressed  the importance of deriving an explicit kinetic equation at order ${ 1/N^{2} }$ in order
to study the collisional relaxation of 1D homogeneous systems in greater detail.

A first step in that direction was made by~\cite{rocha2}.
They started from the equations of the BBGKY hierarchy truncated at order ${ 1/N^{2} }$,
neglected collective effects, and used a computer algebra system  to solve the
truncated hierarchy of equations. However, the form of the collision term that they obtained
was not suitable to study the kinetic equation in detail and solve it. A second step was made
by~\cite{fbcN2} who used a similar procedure and obtained
a more tractable expression  of the kinetic equation at order $1/N^2$. They proved
its well-posedness and established its main properties: conservation laws, $H$-theorem, and
relaxation towards the Boltzmann distribution. They also carried out detailed comparisons with
direct numerical simulations and found a good agreement at sufficiently high temperatures
where collective effects (that are neglected in their kinetic equation) are weak enough.
The kinetic equation at order $1/N^2$ is
fundamentally different from the Boltzmann equation  (or from the related Landau
and Balescu-Lenard equations) because it involves the product of three distribution
functions instead of just two, in line with the fact that the evolution is driven
by three-body correlations instead of two-body correlations. Therefore, it is remarkable
that an $H$-theorem can still be proven in this case by a method which is completely
different from that of Boltzmann. This highlights that the validity of the $H$-theorem
goes beyond the original Boltzmann picture. This also gives a more general justification
(from the kinetic theory angle) of the maximum entropy principle that is used to determine
the statistical equilibrium state of the system.

The kinetic equation derived in~\cite{fbcN2} was restricted to the \HMF\ model,
i.e.\@, to a potential of interaction which involves
only one Fourier mode. In the present paper, we go
beyond these limitations, namely, we generalize the kinetic equation to an arbitrary
potential of interaction. This is an important generalization because it allows us to
treat more general situations of physical interest spanning a wider variety
of long-range interacting potentials.
In the limit where collective effects can be neglected,
i.e.\@, in the limit of dynamically hot systems that only weakly amplify perturbations,
we present a closed and explicit kinetic equation generically describing
the collisional relaxation of the system on ${ N^{2} \td }$ timescales,
as driven by three-body correlations.
Strikingly, for long-range interactions, we show that no further kinetic blocking is possible.
Finally, in addition to exploring the generic properties of this collision operator,
we also quantitatively compare its predictions with direct $N$-body simulations.

The paper is organised as follows.
In Section~\ref{sec:KineticEquation},
we present the kinetic equation describing relaxation at order ${ 1/N^{2} }$,
as given by Eq.~\eqref{kin_eq}.
The detailed procedure used to derive that equation
is described in Appendix~\ref{sec:Derivation},
while the effective calculations were performed
using a computer algebra system
(see Supplemental Material~\citep{MMA}).
In Section~\ref{sec:Properties},
we present the main properties of this kinetic equation,
in particular its conservation laws
and its $H$-theorem.
In Section~\ref{sec:Steady},
we explore in detail the steady states of this kinetic equation,
highlighting in particular that,
as long as the interaction potential is long-range,
${1/N^{2}}$ effects unavoidably lead to the full relaxation of the system
towards the Boltzmann distribution.
In Section~\ref{sec:Wellposedness},
we show that the kinetic equation is well-posed,
i.e.\@, that one can compute explicitly its prediction.
In Section~\ref{sec:Applications},
we illustrate how this equation quantitatively matches
measurements from direct numerical simulations,
for initial conditions dynamically hot enough.
Finally, we conclude in Section~\ref{sec:Conclusions}.

\section{The kinetic equation}
\label{sec:KineticEquation}

We are interested in the long-term dynamics
of a (periodic) 1D long-range interacting system.
We assume that it is composed of $N$ particles
of individual mass ${ \mu = \Mtot / N }$,
with $\Mtot$ the system's total mass.
The canonical phase space coordinates
are denoted by ${ (\theta , v) }$,
with $\theta$ a ${2\pi}$-periodic angle
and $v$ the velocity.
The system's total Hamiltonian then reads
\begin{equation}
H = \frac{1}{2} \sum_{i = 1}^{N} v_{i}^{2} + \mu \sum_{i < j}^{N} U (\theta_{i} , \theta_{j}) ,
\label{Htot}
\end{equation}
where ${ U (\theta_{i} , \theta_{j}) }$ stands
for the considered pairwise interaction potential.
We naturally assume that the potential satisfies the symmetries
${ U (\theta_{i} , \theta_{j}) = U (|\theta_{i} - \theta_{j}|) }$.
As such, it can be expanded in Fourier-space as
\begin{equation}
U (\theta_{1} , \theta_{2}) = - \sum_{k} U_{k} \, \re^{\ri k (\theta_{1} - \theta_{2})} ,
\label{Fourier_Uk}
\end{equation}
where the coefficients, ${ U_{k} \in \mathbb{R} }$,
satisfy the symmetry ${ U_{-k} = U_{k} }$.
In Eq.~\eqref{Fourier_Uk}, we also introduced an overall negative sign,
so that one generically has ${ U_{k} \geq 0 }$ for an attractive potential.

For an homogeneous system,
the instantaneous state of the system is described by
its velocity \DF\@, ${ F (v , t) }$,
which we normalise as ${ \!\int\! \rd \theta \rd v F = \Mtot }$,
with $\Mtot$ the total mass of the system.
To describe the long-term relaxation of the system,
one must characterise the long-term evolution of that \DF\
through a closed self-consistent kinetic equation.

As derived in~\cite{heyvaerts,physicaA} and references therein,
if one limits oneself only to ${1/N}$ effects,
the dynamics of ${ F (v , t) }$ is described
by the homogeneous Balescu-Lenard equation.
With the present notation, it reads
\begin{align}
\frac{\partial F (v)}{\partial t} = & \, 2 \pi^{2} \mu \frac{\partial }{\partial v} \bigg[ \sum_{k} \frac{|k| \, |U_{k}|^{2}}{|\veps_{k} (k \, v)|^{2}} 
\nonumber
\\
\times & \, \!\! \int \!\! \rd v_{1} \, \deltaD (v \!-\! v_{1}) \, \bigg( \frac{\partial }{\partial v} \!-\! \frac{\partial }{\partial v_{1}} \bigg) F (v) F (v_{1}) \bigg] ,
\label{BL_equation}
\end{align}
where the time dependence of the \DFs\ was dropped
to shorten the notations.
In that equation, we also introduced the dielectric function,
${ \veps_{k} (\omega) }$,
whose explicit expression is given in Eq.~\eqref{def_veps}.

Because of the resonance condition, ${ \deltaD (v - v_{1}) }$,
the diffusion flux from the Balescu-Lenard equation~\eqref{BL_equation}
exactly vanishes.
Indeed, only local two-body resonances of the form ${v = v_{1}}$
are permitted,
which, because of the exact local cancellation
of the sum of the drift and diffusion coefficients,
cannot drive any relaxation of the system's mean \DF\@.
One-dimensional homogeneous systems are generically
kinetically blocked w.r.t.\ two-body correlations at order ${1/N}$.
This drastically slows down the system's long-term evolution.
As a consequence, it is only through weaker three-body correlations,
via ${1/N^{2}}$ effects,
that such systems can relax to their thermodynamical equilibrium.
This is the dynamics on which the present paper is focused.

On the one hand, the effective derivation of the system's appropriate kinetic equation
is straightforward,
as the roadmap to follow is systematic.
On the other hand, these calculations rapidly become cumbersome
in practice given the large numbers of terms that one has to deal with.
In addition, to finally reach a simple closed form,
one also has to perform numerous symmetrisations and relabellings.
All in all, to alleviate the technical aspects of these calculations,
we carried out all our derivation using \texttt{Mathematica}
with a code that can be found in the Supplemental Material~\citep{MMA}.
In this paper, we will restrict ourselves to the outline
of the derivation.

The key details of our approach are spelled out in Appendix~\ref{sec:Derivation}.
In a nutshell, the main steps of the derivation are as follows.
(i) First, we derive the usual coupled BBGKY evolution equations
for the one-, two-, and three-body distribution functions,
i.e.\@, the equations that fully encompass the system's
dynamics at order ${1/N^{2}}$.
(ii) Using the cluster expansion~\citep{Balescu1997}, we can rewrite
these evolution equations as coupled equations
for the one-body \DF\@, ${ F (v , t) }$,
and the two- and three-body correlation functions.
At this stage, the evolution equations are still coupled to each other,
but are ordered w.r.t.\ the small parameter ${1/N}$.
(iii) We may then truncate these equations at order ${1/N^{2}}$.
In addition, at this stage, we also neglect the contribution
from collective effects,
assuming that the system is dynamically hot
so that it is not efficient at self-consistently amplifying perturbations.\footnote{
Eventually, this assumption should be lifted to describe colder systems.
}
Another key trick is to split the two-body correlation functions
in two components, respectively associated
with the ${1/N}$ and ${1/N^{2}}$ contributions.
(iv) Finally, having set up a set of four (well-posed)
coupled partial differential equations,
we may solve them explicitly in time.
At that stage, the key assumption is Bogoliubov's ansatz,
i.e.\@, the assumption that the system's mean \DF\
evolves on timescales much longer than its correlation functions.
Following various relabellings, symmetrisations,
and integrations by part,
we finally obtain an explicit and closed expression
for the system's ${1/N^{2}}$ collision operator.
The hardest part of this calculation is the appropriate
use of the resonance conditions to simplify accordingly
the arguments of the functions appearing in the kinetic equation.

All in all, the kinetic equation then reads
\begin{align}
\frac{\partial F (v)}{\partial t} = 2 \pi^{3} \mu^{2} & \, \frac{\partial }{\partial v} \bigg[ 
\!\sum_{k_{1} , k_{2}} \!\! \frac{k_{2}^{2}}{k_{1}^{2} (k_{1} \!+\! k_{2})} \, \mU (k_{1} , k_{2}) \, \mP \!\! \int \!\! \frac{\rd v_{1}}{(v \!-\! v_{1})^{4}}
\nonumber
\\
& \, \times  \!\! \int \!\! \rd v_{2} \, \deltaD \big[ \bk \cdot \bvel \big] \bigg( \bk \cdot \frac{\partial }{\partial \bvel} \bigg) F_{3} (\bvel) \bigg] ,
\label{kin_eq}
\end{align}
where the sum over ${ k_{1}, k_{2} }$ is restricted to the indices
such that $k_{1}$, $k_{2}$, and ${ ( k_{1} \!+\! k_{2} ) }$ are all non-zero.
In Eq.~\eqref{kin_eq}, to shorten the notations, we introduced the velocity vector
${ \bvel = (v , v_{1} , v_{2}) }$, as well as ${ F_{3} (\bvel) = F (v) F (v_{1}) F (v_{2}) }$.
Finally, we introduced the resonance vector
\begin{equation}
\bk = \big( k_{1} \!+\! k_{2} , - k _{1} , - k_{2} \big) 
\label{def_bk}
\end{equation}
as well as the coupling factor
\begin{equation}
\mU (k_{1},k_{2}) \!=\! \big( (k_{1} \!+\! k_{2}) U_{k_{1}} U_{k_{2}} \!-\! k_{1} U_{k_{1} + k_{2}} U_{k_{2}} \!-\! k_{2} U_{k_{1} + k_{2}} U_{k_{1}} \big)^{2} .
\label{def_mU}
\end{equation}
In Eq.~\eqref{kin_eq}, we also introduced
Cauchy's principal value, as $\mP$,
which acts on the integral ${ \!\int\! \rd v_{1} }$.
We postpone to Section~\ref{sec:Wellposedness}
the proof of its well-posedness.

Of course, the similarities between the ${1/N}$
Balescu-Lenard equation~\eqref{BL_equation}
and the present ${1/N^{2}}$ equation are striking.
We emphasise that Eq.~\eqref{kin_eq}
is proportional to ${ \mu^{2} \sim 1/N^{2} }$,
so that it effectively describes a (very) slow
relaxation on ${ N^{2} \td }$ timescales.
In addition, we also note that the collision operator
involves the \DF\ three times,
which stems from the fact that the relaxation
is sourced by three-body correlations.
Such correlations are coupled through a resonance condition
on three distinct velocities,
namely via the factor ${ \deltaD [\bk \cdot \bvel] }$.
This is one of the key changes w.r.t.\ to the ${ 1/N }$ kinetic equation~\eqref{BL_equation},
as the present three-body resonances allow for non-trivial
and non-local kinetic couplings,
driving a non-vanishing overall relaxation.
Equation~\eqref{kin_eq} also differs from Eq.~\eqref{BL_equation} in one other significant manner,
in as much as it does not involve the dielectric function, ${ \veps_{k} (\omega) }$,
since collective effects have been neglected at this stage
(we suggest in footnote~\ref{foot:eps} how collective effects
may be accounted for in Eq.~\eqref{kin_eq}).

Equation~\eqref{kin_eq} is the main result of the paper:
this closed and explicit kinetic equation is the appropriate self-consistent
kinetic equation to describe the long-term evolution of a dynamically
hot one-dimensional homogeneous system,
as driven by ${1/N^{2}}$ effects.
It is quite general
since Eq.~\eqref{kin_eq} applies to any arbitrary long-range interaction potentials,
as defined in Eq.~\eqref{Fourier_Uk}.
Finally, Eq.~\eqref{kin_eq} holds
as long as the system remains linearly Vlasov stable,
to prevent it from being driven to an inhomogeneous state.

\section{Properties}
\label{sec:Properties}

In this section, we explore some of the key properties
of the kinetic equation~\eqref{kin_eq}.

\subsection{Conservation laws}
\label{sec:Conservation}

The kinetic equation~\eqref{kin_eq} satisfies various conservation laws,
in particular the conservation of the total mass, ${ M (t) }$, momentum, ${ P (t) }$,
and energy, ${ E (t) }$.
Ignoring irrelevant prefactors, these quantities are defined as
\begin{align}
M (t) & \, = \!\! \int \!\! \rd v \, F (v , t) ,
\nonumber
\\
P (t) & \, = \!\! \int \!\! \rd v \, v \, F (v , t) ,
\nonumber
\\
E (t) & \, = \!\! \int \!\! \rd v \, \tfrac{1}{2} v^{2} \, F (v , t) .
\label{def_conserved_quantities}
\end{align} 

To recover the conservation of these quantities,
let us first rewrite Eq.~\eqref{kin_eq} as
\begin{equation}
\frac{\partial F (v)}{\partial t} = \frac{\partial }{\partial v} \, \mF (v , t) ,
\label{def_Flux}
\end{equation}
with ${ \mF (v , t) }$ the diffusion flux.
We can then rewrite the time derivatives of Eq.~\eqref{def_conserved_quantities} as
\begin{align}
\frac{\rd M}{\rd t} & \, = \!\! \int \!\! \rd v \, \frac{\partial }{\partial v} \, \mF (v , t) ,
\nonumber
\\
\frac{\rd P}{\rd t} & \, = - \!\! \int \!\! \rd v \, \mF (v , t) ,
\nonumber
\\
\frac{\rd E}{\rd t} & \, = - \!\! \int \!\! \rd v \, v \, \mF (v , t) .
\label{ddt_conserved_quantities}
\end{align}
The conservation of the total mass then follows from the absence
of any boundary contributions, so that one has ${ \rd M / \rd t = 0 }$.

Recovering the conservation of ${ P (t) }$ and ${ E (t) }$ requires a bit more finesse,
as one needs to leverage the symmetry properties of the terms involved.
The main trick is to study the symmetries of the term ${ \!\int\! \rd v \, \mF (v) }$.
One can write
\begin{equation}
\!\! \int \!\! \rd v \, \mF (v) = \sum_{k_{1} , k_{2}} (k_{1} \!+\! k_{2}) \!\! \int \!\! \rd v \rd v_{1} \rd v_{2} \, A_{k_{1}k_{2}} (v,v_{1},v_{2}) ,
\label{sym_Flux}
\end{equation}
where the expression of ${ A_{k_{1}k_{2}} (v,v_{1},v_{2}) }$
follows from Eq.~\eqref{sym_Flux} and reads
\begin{align}
A_{k_{1}k_{2}}(v,v_{1},v_{2}) = & \, 2 \pi^{3} \mu^{2} \frac{k_{2}^{2}}{k_{1}^{2} (k_{1}\!+\!k_{2})^{2}} \mU (k_{1} , k_{2}) \, \mP \!\! \int \!\! \frac{\rd v_{1}}{(v - v_{1})^{4}}
\nonumber
\\
& \, \times \!\! \int \!\! \rd v_{2} \, \deltaD \big[ \bk \cdot \bvel \big] \, \bigg(\! \bk \cdot \frac{\partial }{\partial \bvel} \!\bigg) \, F_{3} (\bvel) .
\label{def_A}
\end{align}

Starting from Eq.~\eqref{sym_Flux}, one can first perform the relabellings ${ \{ v , v_{1} \} \to \{ v_{1} , v_{2} \} }$ and ${ \{ k_{1} , k_{2} \} \to \{ -k_{1}-k_{2} , k_{2} \} }$.
Following these changes, which are more easily performed
using a computer algebra system~\citep{MMA},
Eq.~\eqref{sym_Flux} becomes
\begin{equation}
\!\! \int \!\! \rd v \, \mF (v) = - \sum_{k_{1} , k_{2}} k_{1} \!\! \int \!\! \rd v \rd v_{1} \rd v_{2} \, A_{k_{1}k_{2}} (v , v_{1} , v_{2}) .
\label{sym_Flux_1}
\end{equation}
Similarly, starting once again from Eq.~\eqref{sym_Flux}, one can also perform
the relabellings ${ \{ v , v_{2} \} \to \{ v_{2} , v_{1} \} }$ and ${ \{ k_{1} , k_{2} \} \to \{ -k_{1} , k_{1} + k_{2} \} }$. Following these changes, Eq.~\eqref{sym_Flux} becomes
\begin{equation}
\!\! \int \!\! \rd v \, \mF (v) = - \sum_{k_{1} , k_{2}} k_{2} \!\! \int \!\! \rd v \rd v_{1} \rd v_{2} \, A_{k_{1} k_{2}} (v , v_{1} , v_{2}) .
\label{sym_Flux_2}
\end{equation}

Having obtained the symmetrised expressions from Eqs.~\eqref{sym_Flux},~\eqref{sym_Flux_1},
and~\eqref{sym_Flux_2}, we can now go back to the computation
of the conserved quantities from Eq.~\eqref{ddt_conserved_quantities}.
By adding ${\tfrac{1}{3}}$ of every expression, we obtain
\begin{align}
\frac{\rd P}{\rd t} & \, = - \frac{1}{3} \sum_{k_{1} , k_{2}} \!\! \int \!\! \rd v \rd v_{1} \rd v_{2} \, A_{k_{1}k_{2}} (v , v_{1} , v_{2})
\nonumber
\\
& \;\;\;\;\;\;\;\;\;\;\; \times \bigg\{ (k_{1} \!+\! k_{2}) - k_{1} - k_{2} \bigg\}
\nonumber
\\
& \, = 0 .
\label{calc_dPdt}
\end{align}

We can proceed very similarly for the total energy,
repeating the symmetrisations which were performed to obtain
the various rewritings of the integral of the flux.
Equation~\eqref{ddt_conserved_quantities} becomes
\begin{align}
\frac{\rd E}{\rd t} & \, = - \frac{1}{3} \sum_{k_{1} , k_{2}} \!\! \int \!\! \rd v \rd v_{1} \rd v_{2} \, A_{k_{1}k_{2}}(v,v_{1},v_{2})
\nonumber
\\
& \;\;\;\;\;\;\;\;\;\;\; \times \bigg\{ (k_{1} \!+\! k_{2}) v - k_{1} v_{1} - k_{2} v_{2} \bigg\}
\nonumber
\\
& \, = 0 ,
\label{calc_dEdt}
\end{align}
owing to the presence of the resonance condition ${ \deltaD [(k_{1} \!+\! k_{2}) v \!-\! k_{1} v_{1} \!-\! k_{2} v_{2}] }$ in the expression of ${ A_{k_{1} k_{2}} (v , v_{1}  ,v_{2}) }$ in Eq.~\eqref{def_A}.

\subsection{$H$-theorem}
\label{sec:HTheorem}

Let us define the system's entropy as
\begin{equation}
S (t) = - \!\! \int \!\! \rd v \, s \big( F (v , t) \big) ,
\label{def_S}
\end{equation}
with ${ s (F) = F \, \ln (F) }$ Boltzmann's entropy.
Following the definition from Eq.~\eqref{def_Flux},
the time derivative of Eq.~\eqref{def_S} reads
\begin{equation}
\frac{\rd S}{\rd t} = \!\! \int \!\! \rd v \, \frac{F' (v)}{F (v)} \, \mF (v , t) .
\label{dSdt}
\end{equation}
To show that the system's entropy unavoidably and systematically grows with time,
we use the same approach as in the previous section.
Repeating the symmetrisations which were performed in Eqs.~\eqref{sym_Flux_1}
and~\eqref{sym_Flux_2},
we can rewrite Eq.~\eqref{dSdt} as
\begin{align}
\frac{\rd S}{\rd t} = & \, \frac{1}{3} \sum_{k_{1} , k_{2}} \!\! \int \!\! \rd v \rd v_{1} \rd v_{2} \, A_{k_{1} k_{2}} (v , v_{1} , v_{2})
\nonumber
\\
\times & \, \bigg\{ (k_{1} \!+\! k_{2}) \, \frac{F'(v)}{F (v)} - k_{1} \frac{F'(v_{1})}{F(v_{1})} - k_{2} \frac{F'(v_{2})}{F(v_{2})} \bigg\} .
\label{calc_dSdt}
\end{align}
Luckily, returning to the definition of ${ A_{k_{1} k_{2}} }$ from Eq.~\eqref{def_A},
we note that Eq.~\eqref{calc_dSdt} can be rewritten under the form
\begin{align}
\frac{\rd S}{\rd t} & \, = \frac{2 \pi^{3} \mu^{2}}{3} \sum_{k_{1} , k_{2}} \!\! \int \!\! \rd v \rd v_{1} \rd v_{2} \, \frac{k_{2}^{2}}{k_{1}^{2} (k_{1} \!+\! k_{2})^{2}} \,  \mU (k_{1} , k_{2}) 
\nonumber
\\
& \, \times \mP \bigg(\! \frac{1}{(v - v_{1})^{4}} \!\bigg) \, \frac{\deltaD \big[ \bk \cdot \bvel \big]}{F_{3} (\bvel)}
\nonumber
\\
& \, \times \bigg( (k_{1} \!+\! k_{2}) \frac{F'(v)}{F(v)} - k_{1} \frac{F'(v_{1})}{F(v_{1})} - k_{2} \frac{F'(v_{2})}{F(v_{2})} \bigg)^{2} .
\label{rewrite_dSdt}
\end{align}
As all the terms in these integrals are positive,
in particular the interaction coupling ${ \mU (k_{1} , k_{2}) }$ from Eq.~\eqref{def_mU},
the kinetic equation~\eqref{kin_eq} therefore satisfies an $H$-theorem,
i.e.\@, one has
\begin{equation}
\frac{\rd S}{\rd t} \geq 0 .
\label{HTheorem}
\end{equation}
This is the essential result of the present section,
as we have just proven that the kinetic equation~\eqref{kin_eq}
unavoidably leads to an irreversible relaxation
of the system.
In Section~\ref{sec:Steady}, we will use the expression of the entropy increase
from Eq.~\eqref{rewrite_dSdt} to determine which \DFs\ are the equilibrium
states of the diffusion, i.e.\@, which \DFs\ satisfy ${ \rd S / \rd t = 0 }$.

\subsection{Dimensionless rescaling}
\label{sec:Dimensionless}

We introduce the system's velocity dispersion as
\begin{equation}
\sigma^{2} = \frac{1}{\Mtot} \!\! \int \!\! \rd \theta \rd v \, v^{2} \, F (v) .
\label{def_sigma}
\end{equation}
This entices us then to also introduce the dimensionless velocity, $u$,
and time, $\ot$, as
\begin{equation}
u = \frac{v}{\sigma} \;\;\; ; \;\;\; \ot = \frac{t}{\td} ,
\label{def_u_ot}
\end{equation}
with ${ \td = 1 / \sigma }$ the system's dynamical time.
Similarly, it is natural to introduce the dimensionless \PDF\
\begin{equation}
\oF (u) = \frac{2 \pi \sigma}{\Mtot} \, F (u \sigma) ,
\label{def_PDF}
\end{equation}
which satisfies the normalisation condition ${ \!\int\! \rd u \oF (u) = 1 }$.
We note that this \PDF\ has a (dimensionless) unit velocity dispersion
given by ${ \!\int\! \rd u \, u^{2} \oF(u) = 1 }$.
Finally, we must also introduce a quantity to assess the dynamical temperature
of the system, and the strength of the associated underlying collective effects.
Following Appendix~\ref{sec:LinearTheory},
we define the dimensionless stability parameter
\begin{equation}
Q = \frac{\sigma^{2}}{\Umax \Mtot} , 
\label{def_Q}
\end{equation}
where we introduced ${ \Umax = \max_{k} U_{k} }$.
The larger $Q$, the hotter the system, i.e.\@, the weaker the collective effects.
Given $\Umax$,
we may finally define the dimensionless interaction coefficients
${ \oU_{k} = U_{k} / \Umax }$.

Using these conventions, we can rewrite Eq.~\eqref{kin_eq}
under the dimensionless form
\begin{align}
& \, \frac{\partial \oF (u)}{\partial \ot} \!=\! \frac{\pi}{2} \frac{1}{Q^{4} N^{2}} \frac{\partial }{\partial u} \bigg[  \sum_{k_{1} , k_{2}} \!\! \frac{k_{2}^{2}}{k_{1}^{2} (k_{1} \!+\! k_{2})} \omU (k_{1},k_{2}) \,
\nonumber
\\
& \times \mP \!\! \int \!\! \frac{\rd u_{1}}{(u \!-\! u_{1})^{4}} \!\! \int \!\! \rd u_{2} \bigg\{ \deltaD \big[ \bk \cdot \bu \big] \bigg(\! \bk \cdot \frac{\partial }{\partial \bu} \!\bigg) \, \oF_{3} (\bu) \bigg\} \bigg] ,
\label{kin_eq_ddim}
\end{align}
where the coupling factor ${ \omU (k_{1},k_{2}) }$ naturally follows from Eq.~\eqref{def_mU}
with the replacement ${ U_{k} \to \oU_{k} }$.
Finally, Eq.~\eqref{kin_eq_ddim}
can be rewritten as a continuity equation, reading
\begin{equation}
\frac{\partial \oF (u)}{\partial \ot} = \frac{\pi}{2} \frac{1}{Q^{4}N^{2}} \, \frac{\partial }{\partial u} \big[ \, \omF (u) \big] ,
\label{def_omF}
\end{equation}
where the dimensionless instantaneous flux, ${ \omF (u) }$,
follows from Eq.~\eqref{kin_eq_ddim}.

Equation~\eqref{kin_eq_ddim} is an enlightening rewriting
of the kinetic equation,
as it clearly highlights the expected relaxation time
of a given system.
Assuming that the term within brackets is of order unity,
Eq.~\eqref{kin_eq_ddim} states therefore that the relaxation time, $\tr$,
of the system scales like
\begin{equation}
\tr \simeq Q^{4} N^{2} \td . 
\label{scaling_relaxation_time}
\end{equation}
In particular, we recover that the hotter the system,
the slower the long-term relaxation.
As Eq.~\eqref{kin_eq} was derived while neglecting collective effects,
i.e.\@, in the limit ${ Q \gg 1 }$,
the relaxation will only occur on very very long timescales
because of the factor $Q^{4}$ in Eq.~\eqref{scaling_relaxation_time}.

\section{Steady states}
\label{sec:Steady}
 
In the previous section, we showed that Eq.~\eqref{kin_eq}
satisfies an H-theorem for the Boltzmann entropy.
Let us now explore what are the steady states of that evolution equation,
i.e.\@, the \DFs\ such that Eq.~\eqref{kin_eq} predicts ${ \partial F / \partial t = 0 }$.

 \subsection{Boltzmann distribution}
\label{sec:Boltzmann}

We expect the thermodynamical equilibria originating from relaxation
to take the form of (possibly shifted) homogeneous Boltzmann \DF\ reading
\begin{equation}
\FB (v) = C \, \re^{- \beta (v - v_{0})^{2}} ,
\label{def_Boltzmann}
\end{equation}
with $\beta$ the inverse temperature, and $C$ a normalisation constant.
These \DFs\ maximise the Boltzmann entropy
at fixed mass, momentum, and energy.
It is straightforward to check that such \DFs\ are equilibrium solutions
of the kinetic equation~\eqref{kin_eq}.
Indeed, noting that the vector $\bk$ from Eq.~\eqref{def_bk}
is of zero sum, we can write
\begin{align}
\frac{\partial \FB (v)}{\partial t} \propto \deltaD \big[ \bk \cdot \bvel \big] \, \big( \bk \cdot \bvel \big) = 0 .
\label{vanish_Boltzmann}
\end{align}
This is an important result, as it highlights that homogeneous Boltzmann distributions
are indeed equilibrium solutions of the ${1/N^{2}}$ kinetic equation~\eqref{kin_eq}.
In the coming sections, thanks to the H-theorem,
we will strengthen this result by showing that homogeneous Boltzmann \DFs\
are in fact the only equilibrium solutions of the present kinetic equation,
whatever the considered long-range interacting potential.

\subsection{Constraint from the H-theorem}
\label{sec:ConstraintHTheorem}
 
Following the computation of ${ \rd S / \rd t }$ in Eq.~\eqref{HTheorem},
we can now determine what are the most generic steady states
of the kinetic equation~\eqref{kin_eq}.
Assuming that there exists $\bk$ such that ${ \mU (k_{1} , k_{2}) \neq 0 }$,
and introducing the function ${ G(v) = F'(v)/F(v) }$,
a \DF\ nullifies the rate of entropy if it satisfies
\begin{equation}
\forall v_{1} , v_{2} : \, G \bigg( \frac{k_{1} v_{1} + k_{2} v_{2}}{k_{1} + k_{2}} \bigg) = \frac{k_{1} G (v_{1}) + k_{2} G (v_{2})}{k_{1} + k_{2}} .
\label{condition_dSdt}
\end{equation}
In essence, Eq.~\eqref{condition_dSdt} takes the form of a weighted mean,
with weights $k_{1}$ and $k_{2}$.
As a consequence, for Eq.~\eqref{condition_dSdt} to be satisfied for all $v_{1}$ and $v_{2}$,
the function ${ v \mapsto G(v) }$ must necessarily be a line,
i.e.\@, one must have
\begin{equation}
G (v) = - 2 \beta (v - v_{0}) ,
\label{line_condition}
\end{equation}
with $\beta$ positive to satisfy the contraint
${ \!\int\! \rd \theta \rd v F (v) \!=\! \Mtot }$.
Recalling that ${ G (v) \!=\! F'(v) / F(v) }$, Eq.~\eqref{line_condition}
immediately integrates to the (shifted) homogeneous Boltzmann \DF\ from Eq.~\eqref{def_Boltzmann},
which is already a known equilibrium state,
as detailed in Eq.~\eqref{vanish_Boltzmann}.

As a conclusion, provided that there exists at least one ${ \mU (k_{1},k_{2}) \neq 0 }$,
the only equilibrium \DFs\ of the kinetic equation~\eqref{kin_eq}
are the (shifted) homogeneous Boltzmann distributions.
This is an important result.
Indeed, while any stable \DF\@, ${ F(v) }$,
is systematically an equilibrium distribution for the ${1/N}$ dynamics
of long-range interacting homogeneous systems,
only homogeneous Boltzmann \DFs\ are equilibrium distributions
for the underlying ${1/N^{2}}$ dynamics.
Since the entropy is bounded from above,
the system necessarily relaxes towards these \DFs\@.

\subsection{Constraint from the interaction potential}
\label{sec:ConstraintPotential}

In the previous discussion, in order to recover the unicity
of the steady states, we had to assume that there existed
at least one ${ \mU (k_{1} , k_{2}) \neq 0 }$.
Let us now briefly explore the implications of that assumption.

One can note that the flux from Eq.~\eqref{kin_eq}
exactly vanishes if, for all ${ k_{1} , k_{2} > 0 }$, one has
\begin{equation}
(k_{1} \!+\! k_{2}) U_{k_{1}} U_{k_{2}} = k_{1} U_{k_{1} + k_{2}} U_{k_{2}} + k_{2} U_{k_{1} + k_{2}} U_{k_{1}} .
\label{constraint_U}
\end{equation}
An interaction potential that systematically satisfies
the constraint from Eq.~\eqref{constraint_U} leads to a vanishing flux.

Let us therefore consider ${ n \!>\! 0 }$ as the smallest index such that ${ U_{n} \!\neq\! 0 }$.
Considering the case ${ (k_{1} , k_{2}) \!=\! (n,n) }$ in Eq.~\eqref{constraint_U},
we obtain ${ U_{2n} \!=\! U_{n} }$.
Repeating the operation with ${ (k_{1} , k_{2}) \!=\! (n , 2n) }$,
we can subsequently obtain ${ U_{3n} \!=\! U_{2n} \!=\! U_{n} }$.
Proceeding by recurrence with ${ (k_{1} , k_{2}) \!=\! (n , k \!\times\! n) }$,
we can finally conclude that ${ U_{n} \!=\! U_{2n} \!=\! ... \!=\! U_{k \times n} \!=\! ... }$.
In a similar fashion, let us consider a number ${ \np \!>\! 0 }$,
with ${ \np \!=\! k \!\times\! n + d }$ and ${ 0 \!<\! d \!<\! n }$.
By considering the pair ${ (k_{1} , k_{2}) = (k \!\times\! n , d) }$ in Eq.~\eqref{constraint_U},
we conclude that ${ U_{\np} \!=\! 0 }$,
where we used the fact that ${ U_{d} \!=\! 0 }$ by assumption
since ${ d \!<\! n }$.

To summarise, the only non-trivial solutions to the constraint
from Eq.~\eqref{constraint_U} are indexed by an integer ${ n > 0 }$,
and read
\begin{equation}
U_{k} = 
\begin{cases}
\displaystyle 0 & \, \text{ if } \;\;\; k = 0 ,
\\
\displaystyle U_{0} & \text{ if } \;\;\; |k| > 0 \text{ and } k \equiv 0 \;\; \mathrm{mod} \; n ,
\\
\displaystyle 0 & \text{ otherwise}.
\end{cases}
\label{solution_U}
\end{equation}

Thankfully, once the Fourier transform of the potential has been characterised
via Eq.~\eqref{solution_U},
one can straightforwardly compute its expression in $\theta$-space.
It reads
\begin{equation}
U (\theta) = U_{0} \bigg( 1 - \frac{1}{n} \sum_{k = 0}^{n-1} \deltaD \bigg[ \theta - k \frac{\pi}{n} \bigg] \bigg) .
\label{exp_U_constraint}
\end{equation}
The generic class of potentials from Eq.~\eqref{exp_U_constraint} are the only potentials
for which the flux from Eq.~\eqref{kin_eq} systematically vanishes,
whatever the \DF\@.
Because Eq.~\eqref{exp_U_constraint} involves Dirac deltas,
it does not correspond to a long-range interaction,
but rather to an exactly local interaction.
Of particular interest is the case ${ n = 1 }$, which leads to the simple Dirac interaction,
${ U (\theta) = U_{0} \big( 1 - \deltaD (\theta) \big) }$.
The dynamics driven by this potential is identical to the dynamics
of pointwise marbles on the circle that would undergo hard collisions.
In such a system, when two marbles collide, they exactly reverse their velocity:
this cannot induce any relaxation of the system's overall \DF\@, ${ F (v) }$.
Hence, we have shown that systems with local interactions generically
undergo a kinetic blocking also for the ${1/N^{2}}$ dynamics.
Following Eq.~\eqref{exp_U_constraint},
we have also shown that there exist no long-range interaction potentials
for which one can devise a kinetic blocking of the ${1/N^{2}}$ kinetic blocking.
This is an important result.
As soon as the considered interaction potential, ${ U (\theta) }$, is not exactly local,
the homogeneous Boltzmann \DFs\ from Eq.~\eqref{def_Boltzmann}
are the only equilibrium states of the kinetic equation~\eqref{kin_eq}.
Furthermore, the $H$-theorem guarantees that these equilibrium states
are reached for ${ t \to + \infty }$ (in practice for ${ t \gtrsim N^{2} \td }$).
Three-point correlations are always able to induce relaxation
for long-range interacting homogeneous 1D systems.

\section{Well-posedness}
\label{sec:Wellposedness}

As a result of the presence of a high-order resonance denominator
in Eq.~\eqref{kin_eq}, it is not obvious a priori that this equation is well-posed,
i.e.\@, that there are no divergences when ${ v_{1} \to v }$.
We will now show that Eq.~\eqref{kin_eq} can be rewritten
under an alternative form
allowing for the principal value to be computed.
The required symmetrisations and relabellings are in fact quite subtle.

Let us first rewrite Eq.~\eqref{kin_eq} under a form that better captures
its resonant structure.
We define the set of fundamental resonances as
\begin{equation}
\big\{ (k , \kp) \; \big| \; 0 < k , \kp \big\} .
\label{def_bbF}
\end{equation}
Then, for a given fundamental resonance, ${ (k , \kp) }$,
there exists a set of resonance pairs, ${ (k_{1} , k_{2}) }$,
associated with the resonance numbers appearing in the sum of Eq.~\eqref{kin_eq}.
This set reads
\begin{align}
\mR (k , k') = \big\{ & \, (k , \kp) , (k + \kp , - k) , (k , -k - \kp)
\nonumber
\\
& \, (\kp , k) , (k + \kp , - \kp) , (\kp , -k - \kp) \big\} ,
\label{def_R}
\end{align}
noting that even for ${ k = \kp }$,
this set still contains six elements.
We also note that all the elements ${ (k_{1} , k_{2}) }$ in ${ \mR (k , \kp) }$
are such that ${ k_{1} > 0 }$.

Following these definitions, we can rewrite Eq.~\eqref{kin_eq} as
\begin{align}
\frac{\partial F (v)}{\partial t} = 2 \pi^{3} & \, \mu^{2} \frac{\partial }{\partial v} \bigg[ \sum_{k , \kp > 0} \mU (k , \kp) \, \mP \!\! \int \!\! \frac{\rd v_{1}}{(v - v_{1})^{4}} 
\nonumber
\\
& \, \times \sum_{(k_{1} , k_{2}) \in \mR (k , \kp)} \frac{k_{2}^{2}}{k_{1}^{2} (k_{1} + k_{2})}
\nonumber
\\
& \, \times \!\! \int \!\! \rd v_{2} \, \deltaD \big[ \bk \cdot \bvel \big] \, \bigg( \bk \cdot \frac{\partial }{\partial \bvel} \bigg) \, F_{3} (\bvel) \bigg] .
\label{rewrite_kin_eq}
\end{align}
To obtain the correct prefactor in Eq.~\eqref{rewrite_kin_eq},
we noted that the resonance pairs ${ (k_{1} , k_{2}) }$ and ${ (- k_{1} , -k_{2}) }$
have the exact same contribution to the flux,
hence the restriction to the sole elements with ${ k_{1} > 0 }$ in ${ \mR (k , \kp) }$,
in Eq.~\eqref{def_R}.
We also note that the fundamental resonances ${ (k , \kp) }$ and ${ (\kp , k) }$
have the exact same contribution to the overall diffusion flux.
All in all, these two remarks justify why Eqs.~\eqref{kin_eq} and~\eqref{rewrite_kin_eq}
share the exact same prefactor.

The main benefit from Eq.~\eqref{rewrite_kin_eq}
is that all the resonance pairs ${ (k_{1} , k_{2}) }$ associated
with the same fundamental resonance ${ (k , \kp) }$
share the exact same coupling factor, ${ \mU (k , \kp) }$,
as already introduced in Eq.~\eqref{def_mU}.
In order to further shorten the notations, we can subsequently rewrite Eq.~\eqref{rewrite_kin_eq}
as
\begin{equation}
\frac{\partial F (v)}{\partial t} = 2 \pi^{3} \mu^{2} \frac{\partial }{\partial v} \bigg[ \sum_{k , \kp > 0} \, \mU (k , \kp) \, \mF_{(k , \kp)} (v) \bigg] ,
\label{rewrite_kin_eq_2}
\end{equation}
where ${ \mF_{(k , \kp)} (v) }$ stands for the flux generated by the fundamental resonance
${ (k , \kp) }$ and reads
\begin{equation}
\mF_{(k , \kp)} (v) = \mP \!\! \int \!\! \frac{\rd v_{1}}{(v - v_{1})^{4}} \!\!\!\! \sum_{(k_{1} , k_{2}) \in \mR (k , \kp)} \!\!\!\! \mC_{(k_{1} , k_{2})} (v , v_{1}) .
\label{def_mF_fund}
\end{equation}
Here, ${ \mC_{(k_{1} , k_{2})} (v , v_{1}) }$ stands for the contribution
from the resonance pair ${ (k_{1} , k_{2}) }$ associated with the fundamental resonance ${ (k , \kp) }$.
Its expression naturally follows from Eq.~\eqref{rewrite_kin_eq}
and, given Eq.~\eqref{def_bk},
reads
\begin{equation}
\mC_{(k_{1} , k_{2})} (v , v_{1}) \!=\! \frac{k_{2}^{2}}{k_{1}^{2} (k_{1} \!+\! k_{2})} \!\! \int \!\! \rd v_{2} \, \deltaD \big[ \bk \cdot \bvel \big] \, \bigg(\! \bk \cdot \frac{\partial }{\partial \bvel} \!\bigg) \, F_{3} (\bvel) .
\label{def_mC}
\end{equation}

The main step to obtain a well-posed writing for the kinetic equation
is to note that in Eq.~\eqref{rewrite_kin_eq},
we perform an integration w.r.t.\ ${ \rd v_{1} \rd v_{2} }$.
As a consequence, we can propose an alternative for ${ \mC_{(k_{1} , k_{2})} (v , v_{1}) }$
by performing the relabelling ${ v_{1} \!\leftrightarrow\! v_{2} }$.
Following that relabelling (see~\cite{MMA}),
we obtain an alternative writing for ${ \mC_{(k_{1} , k_{2})} (v , v_{1}) }$ reading
\begin{equation}
\omC_{(k_{1} , k_{2})} (v , v_{1}) = \frac{k_{1}^{2}}{k_{2}^{2} (k_{1} + k_{2})} \!\! \int \!\! \rd v_{2} \, \deltaD \big[ \obk \cdot \bvel \big] \, \bigg(\! \obk \cdot \frac{\partial }{\partial \bvel} \!\bigg) F_{3} (\bvel) ,
\label{def_omC}
\end{equation}
where, similarly to Eq.~\eqref{def_bk}, we introduced the vector
\begin{equation}
\obk = \big( k_{1} \!+\! k_{2} , - k_{2} , - k_{1} \big) ,
\label{def_obk}
\end{equation}
where $k_{1}$ and $k_{2}$ are flipped w.r.t.\ Eq.~\eqref{def_bk}.
To obtain Eq.~\eqref{def_omC}, we used the presence of the Dirac delta
to make sure that the principal value appears under the form ${ \mP (1 / (v - v_{1})^{4}) }$.
The main changes between Eqs.~\eqref{def_mC} and~\eqref{def_omC}
is a change in the prefactor and the resonance vector to consider.
At this stage, thanks to this alternative writing,
we now have at our disposal all the needed ingredients
to write a well-posed expression for the flux ${ \mF_{(k , \kp)} (v) }$.

The next trick will be to use Eq.~\eqref{def_omC}
on a well chosen subset of the resonance pairs ${ (k_{1} , k_{2}) }$ associated
with a given fundamental resonance ${ ( k , \kp) }$.
Naively, following Eq.~\eqref{def_mF_fund}
and its definition of the resonance pairs,
the flux contribution, ${ \mF_{(k , \kp)} (v) }$, from the fundamental resonance
would read
\begin{align}
\mF_{(k , \kp)} (v) \!=\! \mP \!\! \int \!\! & \frac{\rd v_{1}}{(v - v_{1})^{4}} \bigg\{ \mC_{(k , \kp)} \!+\! \mC_{(k + \kp , - k)} \!+\! \mC_{(k , -k - \kp)}
\nonumber
\\
+ & \,  \big( k \leftrightarrow \kp \big) \bigg\} ,
\label{naive_mF}
\end{align}
where, for clarity,
we dropped the argument ${ (v , v_{1}) }$ from the flux contribution.
Unfortunately, such a writing is still ill-posed, as one can check that the integrand,
for ${ v_{1} = v + \delta v }$, behaves like ${ (\delta v)^{2} }$,
which does not allow for a meaningful computation
of the principal value ${ \mP (1/(\delta v)^{4}) }$.

Let us therefore rewrite Eq.~\eqref{naive_mF} as
\begin{align}
\mF_{(k , \kp)} (v) & \, = \mP \!\! \int \!\! \frac{\rd v_{1}}{(v - v_{1})^{4}} \bigg\{ \mC_{(k , \kp)} + \mC_{(k + \kp , - k)}
\nonumber
\\
+ & \, \bigg( \frac{(k - \kp)^{2}}{k^{2} + \kpsq} \, \mC_{(k , - k - \kp)} + \frac{2 k \kp}{k^{2} + \kpsq} \, \omC_{(k , - k - \kp)} \bigg) 
\nonumber
\\
+ & \, \big( k \leftrightarrow \kp \big) \bigg\} .
\label{smart_mF}
\end{align}
To go from Eq.~\eqref{naive_mF} to Eq.~\eqref{smart_mF},
we replaced ${ \mC_{(k , - k - \kp)} }$ by a weighted average of itself
and its alternative writing ${ \omC_{(k , - k - \kp)} }$.
Such a weighted average is legitimate since we have ${ k^{2} \!+\! \kpsq \!>\! 0 }$,
and the sum of the weights appearing in Eq.~\eqref{smart_mF} satisfies
\begin{equation}
\frac{(k - \kp)^{2}}{k^{2} + \kpsq} + \frac{2 k \kp}{k^{2} + \kpsq} = 1 .
\label{sum_weights}
\end{equation}
When written explicitly, the expression of the flux from Eq.~\eqref{def_mF_fund}
stemming from Eq.~\eqref{smart_mF} reads
\begin{align}
& \, \mF_{(k , \kp)} (v) = \mP \!\! \int \!\! \frac{\rd v_{1}}{(v \!-\! v_{1})^{4}} \!\! \int \!\! \rd v_{2} \, \bigg\{
\nonumber
\\
& \, \frac{\kpsq}{k^{2} (k \!+\! \kp)} \deltaD \big[ \bk \!\cdot\! \bvel \big] \bigg(\! \bk \!\cdot\! \frac{\partial }{\partial \bvel} \!\bigg) F_{3} (\bvel) \bigg|_{\bk = (k + \kp , -k , - \kp)}
\nonumber
\\
+ & \, \frac{\kpsq}{k (k^{2} \!+\! \kpsq)} \deltaD \big[ \bk \!\cdot\! \bvel \big] \bigg(\! \bk \!\cdot\! \frac{\partial }{\partial \bvel} \!\bigg) F_{3} (\bvel) \bigg|_{\bk = (k , - k - \kp , \kp)}
\nonumber
\\
+ & \, \frac{(k \!-\! \kp)^{2} (k \!+\! \kp)^{2}}{k \kpsq (k^{2} \!+\! \kpsq)} \deltaD \big[ \bk \!\cdot\! \bvel \big] \bigg(\! \bk \!\cdot\! \frac{\partial }{\partial \bvel} \!\bigg) F_{3} (\bvel) \bigg|_{\bk = (k , \kp , - k - \kp)}
\nonumber
\\
+ & \, \big( k \leftrightarrow \kp \big) \bigg\} ,
\label{smart_mF_fund}
\end{align}
where we recall that the symmetrisation ${ ( k \leftrightarrow \kp) }$
also applies in the case ${ k = \kp }$.
The crucial gain from Eq.~\eqref{smart_mF_fund} is that the principal value
therein is now well-posed.
Indeed, we may rewrite Eq.~\eqref{smart_mF_fund} as
\begin{equation}
\mF_{(k , \kp)} (v) = \mP \!\! \int \!\! \frac{\rd v_{1}}{(v - v_{1})^{4}} \, K (v , v_{1}) .
\label{def_K_wellposed}
\end{equation}
Assuming that ${ F (v) }$ is a smooth function,
one can then perform a Taylor development of ${ K (v , v \!+\! \delta v) }$
for ${ \delta v \to 0 }$.
One gets (see~\cite{MMA})
\begin{equation}
K (v , v \!+\! \delta v) = K_{3} (v) \, ( \delta v )^{3} + \mO \big( (\delta v)^{4} \big) .
\label{DL_K}
\end{equation}
In the vicinity of ${ v_{1} \to v }$,
Eq.~\eqref{def_K_wellposed} then takes the form
\begin{equation}
\mP \!\! \int \!\! \frac{\rd v}{(v - v_{1})^{4}} \, K (v , v_{1}) \sim \mP \!\! \int \!\! \rd \delta v \, \bigg( \frac{K_{3}}{\delta v} + \mO (1) \bigg) ,
\label{K_wellposed}
\end{equation}
which is a well-posed principal value.
As a conclusion, Eq.~\eqref{smart_mF_fund}
is therefore the form that one must use to explicitly estimate
the diffusion flux, as presented in section~\ref{sec:Applications}.
Another benefit from the writing of Eq.~\eqref{smart_mF_fund}
is that it is the one that allows for an immediate and exact recovery of the
${1/N^{2}}$ kinetic equation already presented in~\cite{fbcN2},
in the (simpler) case of the \HMF\ model,
i.e.\@, a model where only the harmonics ${ k \!=\! 1 }$
is present in the interaction potential.

\textit{Remark} --- Another interest of Eq.~\eqref{rewrite_kin_eq}
is to better understand the scaling of the resonant contributions
for ${k , \kp \to 0 }$ in infinite systems.
This is of particular importance for the Coulombian interaction,
driving the evolution of 1D plasmas~\citep{dawson2}.
In that case, one has ${ U_{k} \propto 1/k^{2} }$.
In the limit where ${k,\kp}$ become continuous variables,
we can transfrom ${ \sum_{k , \kp}\! }$
into ${ \!\int\! \rd k \rd \kp }$,
and we obtain, from Eq.~\eqref{rewrite_kin_eq_2},
the asymptotic behaviour
${ \sim \!\int\! \rd k \rd \kp /k^{\prime\prime 7} }$,
where ${ k^{\prime\prime} }$ is an approximate notation
to refer to either $k$, $\kp$, or ${ (k \!+\! \kp) }$.
While convergent on small scales,
this integral diverges on large scales
(i.e.\@, for ${ k , \kp \to 0 }$).
Such a divergence is, of course, reminiscent of the large-scale divergence
${ \!\int\! \rd k / k^{3} }$
that already appears in the ${1/N}$ Landau equation in 1D,
i.e.\@, the limit ${ \veps_{k} (\omega) \to 1 }$ of Eq.~\eqref{BL_equation}.
The present divergence stems from our neglect of collective effects,
i.e.\@, of the dielectric function ${ \veps_{k} (\omega) }$.
Indeed, on large scales this polarisation leads to Debye shielding,
which ensures the convergence of the collision operator on large scales.

\section{Numerical validation}
\label{sec:Applications}

In order to test the prediction of the kinetic equation~\eqref{kin_eq}
on a full $N$-body system, we carry out numerical simulations
of the (softened) Ring model on the circle
(see, e.g.\@,~\cite{Sota2001,RochaFilho2014}).
This model is characterised by the Hamiltonian
\begin{equation}
H = \frac{1}{2} \sum_{i = 1}^{N} v_{i}^{2} - \sum_{i < j}^{N} \mu \frac{1}{\sqrt{1 - \cos (\theta_{i} - \theta_{j}) + \epsilon}} ,
\label{def_H_ring}
\end{equation}
where $\epsilon$ is a given softening length.
The choice of the Hamiltonian is somewhat ad hoc here,
and was guided by its anharmonicity.

The main difficulty with such a numerical exploration
is that the Hamiltonian from Eq.~\eqref{def_H_ring}
is associated with a fully coupled $N$-body system.
As a consequence, the computational complexity of its time integration
scales like ${ \mO (N^{2}) }$.
This is much more costly than the \HMF\ model investigated in~\citep{fbcN2},
whose dynamics can be integrated in ${ \mO (N) }$,
owing to the presence of globally shared magnetisations.
Simulations are made even harder here because of the need
to consider initial conditions with ${ Q \gg 1 }$,
as Eq.~\eqref{kin_eq} only applies in the limit of dynamically hot initial conditions.
Following the scaling from Eq.~\eqref{scaling_relaxation_time},
relaxation will only occur on very long timescales,
requiring for the simulations to be integrated up to very late times.
Finally, as the potential from Eq.~\eqref{def_H_ring} is quite sharp,
it asks for small integration timesteps, which further increases
the difficulty of reaching very late times.
In order to accelerate our simulations,
we performed them on \GPUs\@.
We give the full details of our numerical setup
in Appendix~\ref{sec:NumericalSimulations}.

In Fig.~\ref{fig:Flux},
we illustrate the initial dimensionless flux,
${ \omF (u , t = 0) }$,
as defined in Eq.~\eqref{def_omF}.
\begin{figure}[htbp!]
\begin{center}
\includegraphics[width=0.48\textwidth]{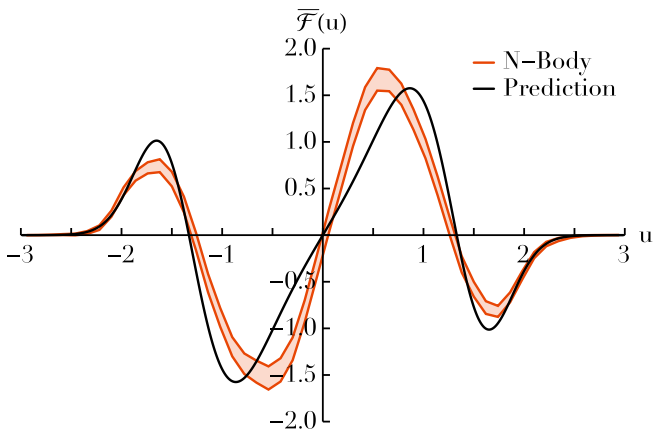}
\caption{Illustration of the dimensionless flux, ${ \omF (u , t \!=\! 0) }$,
as defined in Eq.~\eqref{def_omF}
for the non-Gaussian \PDF\ from Eq.~\eqref{def_nonGaussian}
with ${ \alpha = 4 }$,
and an initial velocity dispersion given by ${ \sigma = 3 }$.
We recover a fairly good quantitative agreement between
the measurements from direct $N$-body simulations
(with the associated errors)
and the prediction of the present kinetic theory
(computed with ${ \kmax = 40 }$, see Fig.~\ref{fig:Predkmax}),
given that the latter is for hot systems only.
Detailed parameters for these runs 
(${ N \!=\! 1024 }$, ${\epsilon \!=\! 0.01}$, and ${ Q \!=\! 9.75 }$)
are spelled out in Appendix~\ref{sec:NumericalSimulations}.
\label{fig:Flux}}
\end{center}
\end{figure}
In that figure, we compare direct measurements
from $N$-body simulations
(following Appendix~\ref{sec:NumericalSimulations})
with the prediction
from the kinetic equation~\eqref{kin_eq}
(using the well-posed rewriting from Eq.~\eqref{smart_mF_fund}).
This figure shows a good quantitative agreement
between the measured and the predicted fluxes.
There are (at least) four possible origins
for the slight mismatch observed in that figure.
(i) There could be some remaining contributions
stemming from collective effects,
still present here for the value ${ Q \simeq 9.75 }$.
(ii) There could be some non-vanishing contributions from the source term
in ${ G_{2}^{(1)} \!\times\! G_{2}^{(1)} }$ that was
neglected in Appendix~\ref{sec:Derivation}
when truncating the BBGKY evolution equations.
(iii) Even with ${ \epsilon = 0.01 }$,
the ring model from Eq.~\eqref{def_H_ring}
still corresponds to a quite hard and local interaction.
As a result, the observed relaxation could
still be partially driven by localised encounters~\citep{Chaffi2017}.
(iv) Finally, one cannot rule out that the numerical simulations
could be partially flawed on such long integration times.

\section{Conclusions}
\label{sec:Conclusions}

This paper presented the closed and explicit kinetic equation
of discrete one-dimensional homogeneous long-range interacting systems
with arbitrary pairwise couplings.
This theory generalises the 
Landau kinetic equation for systems where the ${ 1/N }$ relaxation is blocked by symmetry
and clarifies how three-body correlations
can still drive very-long-term evolutions.
This kinetic equation satisfies mass, momentum, energy conservation,
and an $H$-theorem ensuring relaxation towards the Boltzmann equilibrium.
Provided that the interaction is long-range,
this equation cannot suffer any further kinetic blocking.
As such, Eq.~\eqref{kin_eq} 
represents the ultimate relaxation equation for  classes of hot enough  systems.
Conversely, we have shown that strictly local interactions are kinetically blocked. 
We demonstrated why Eq.~\eqref{kin_eq} is always well-posed,
in spite of the appearance of a fourth order principal value.
We illustrated how this equation quantitatively matches measurements from direct
$N$-body simulations with an anharmonic interaction potential given by Eq.~\eqref{def_H_ring}.
As expected, the much weaker ${ 1/N^{2} }$ interaction leads to a much slower relaxation
requiring very long-term integrations which we carried on \GPUs\@.
The {\sc CUDA} code for these simulations is available on request.

Beyond the scope of this paper, it would clearly be of interest to generalise
Eq.~\eqref{kin_eq} to colder configurations, by taking into account collective polarisations.\footnote{
It is likely that accounting for collective effects
in the derivation of the kinetic equation will ``simply''
amount to dressing the interaction potential,
e.g.\@, making the replacement ${U_{k_{1}} \!\to\! U_{k_{1}}/\veps_{k_{1}} (k_{1} v_{1}) }$
in Eq.~\eqref{def_mU}.\label{foot:eps}
}
In particular, such a generalisation should cure
the large-scale divergence of Eq.~\eqref{kin_eq}
that appears for the Coulombian interaction
(see Section~\ref{sec:Wellposedness}).
Similarly, the present theory could also be expanded
to account for the source term in ${ G_{2}^{(1)} \!\times\! G_{2}^{(1)} }$
(see Appendix~\ref{sec:Derivation}),
that leads to higher order terms in the \DF\@.
Finally, one should also investigate the case of 1D inhomogeneous systems
with monotonic frequency profiles,
that can also suffer from kinetic
blockings~(see, e.g.\@,~\cite{fbc}).
Once these goals are reached,
the kinetic theory of 1D discrete homogeneous long-range
interacting systems will be completed.

\begin{acknowledgments}
This work is partially supported by grant Segal ANR-19-CE31-0017
of the French Agence Nationale de la Recherche: http://www.secular-evolution.org.
We thank St\'ephane Rouberol for the smooth running
of the \GPUs\ on the Horizon cluster,
where the simulations were performed.
\end{acknowledgments}

\appendix

\section{Deriving the kinetic equation}
\label{sec:Derivation}

In this Appendix, we detail the key steps in the derivation
of the kinetic equation~\eqref{kin_eq}.
Notations and normalisations are the sames
as the ones used in~\cite{fbcN2}.

\subsection{BBGKY hierarchy}
\label{sec:BBGKY}

The system is composed of $N$ identical particles
of individual mass ${ \mu = \Mtot/N }$.
We write the phase space coordinates as ${ \bw = (\theta , v) }$.
The instantaneous state of the system is
characterised by its $N$-body \PDF\@,
${ P_{N} (\bw_{1} , ... , \bw_{N} , t) }$,
normalised as ${ \!\int\! \rd \bw_{1} ... \rd \bw_{N} P_{N} = 1 }$,
and assumed to be symmetric w.r.t.\ any permutation of the particles.
This \PDF\ evolves according to Liouville's equation
\begin{equation}
\frac{\partial P_{N}}{\partial t} + \big[ P_{N} , H_{N} \big]_{N} = 0 ,
\label{Liouville_equation}
\end{equation}
where the full $N$-body Hamiltonian reads
\begin{equation}
H_{N} (\bw_{1} , ... , \bw_{N}) = \frac{1}{2} \sum_{i = 1}^{N} v_{i}^{2} + \mu \sum_{i < j}^{N} U (\theta_{i} - \theta_{j}) ,
\label{def_HN}
\end{equation}
with ${ U (\theta_{i} - \theta_{j}) }$ the considered pairwise interaction.
Equation~\eqref{Liouville_equation} also involves the Poisson bracket
over $N$ particles, that is defined with the convention
\begin{equation}
\big[ P_{N} , H_{N} \big]_{N} = \sum_{i = 1}^{N} \bigg( \frac{\partial P_{N}}{\partial \theta_{i}} \frac{\partial H_{N}}{\partial v_{i}} - \frac{\partial P_{N}}{\partial v_{i}} \frac{\partial H_{N}}{\partial \theta_{i}} \bigg) .
\label{def_Poisson}
\end{equation}

In order to better capture the statistical structure of Eq.~\eqref{Liouville_equation},
we introduce the reduced \DFs\@, $F_{n}$, defined as
\begin{equation}
F_{n} (\bw_{1} , ... , \bw_{n} , t) = \mu^{n} \frac{N!}{(N - n)!} \!\! \int \!\! \rd \bw_{n+1} ... \rd \bw_{N} \, P_{N} .
\label{def_Fn}
\end{equation}
With such a choice, we highlight that one has ${ \!\int\! \rd \bw F_{1} (\bw) = \Mtot }$,
so that ${ F_{1} \sim 1 }$ w.r.t.\ $N$ the total number of particles.
The definition from Eq.~\eqref{Liouville_equation} allows us then
to obtain the BBGKY hierarchy as
\begin{equation}
\frac{\partial F_{n}}{\partial t} + \big[ F_{n} , H_{n} \big]_{n} + \!\! \int \!\! \rd \bw_{n+1} \big[ F_{n+1} , \delta H_{n+1} \big]_{n} = 0 ,
\label{BBGKY}
\end{equation}
where we introduced ${ \delta H_{n+1} }$ as the specific interaction energy
of the $(n+1)^{\mathrm{th}}$ particle with the $n$ first.
More precisely, it reads
\begin{equation}
\delta H_{n+1} (\bw_{1} , ... , \bw_{n+1}) = \sum_{i = 1}^{N} U (\theta_{i} - \theta_{n+1}) .
\label{deltaH}
\end{equation}
The first three equations of the BBGKY hierarchy,
i.e.\@, the evolution equations for $F_{1}$, $F_{2}$, and $F_{3}$
are the starting points to derive the kinetic equation.\footnote{
The three-body reduced \DF\@, ${ F_{3} (\bw_{1} , \bw_{2} , \bw_{3}) }$,
should not be confused
with the shortened notation, ${ F_{3} (\bvel) }$,
introduced in Eq.~\eqref{kin_eq}.
}

\subsection{Cluster expansion}
\label{sec:Cluster}

In order to perform a perturbative expansion of the evolution equations,
the next stage of the calculation is to introduce the cluster expansion
of the \DFs\@, following the same normalisation as in~\cite{fbcN2}.

As an example, we introduce the two-body correlation function as
\begin{equation}
F_{2} (\bw_{1} , \bw_{2}) = F_{1} (\bw_{1}) \, F_{1} (\bw_{2}) + G_{2} (\bw_{1} , \bw_{2}) .
\label{def_G2}
\end{equation}
Similar definitions are introduced for the three-body and four-body correlations functions,
${ G_{3} }$ and $G_{4}$.
We do not repeat their definitions here, but refer to Appendix~{B} of~\cite{fbcN2}.

In order to simplify the notations, we now write the one-body \DF\
as ${ F = F_{1} }$.
The dynamical quantities at our disposal then satisfy
the following scalings w.r.t.\ $N$: ${ F \!\sim\! 1 }$, ${ G_{2} \!\sim\! 1/N }$,
${ G_{3} \!\sim\! 1/N^{2} }$, and ${ G_{4} \!\sim\! 1/N^{3} }$.
As such, there are appropriate functions
to perform perturbative expansions w.r.t.\ $N$.

The next step of the calculation is to inject this cluster expansion
into the three first equations of the BBGKY hiearchy,
as given by Eq.~\eqref{BBGKY},
so as to obtain evolution equations for ${ \partial F / \partial t }$,
${ \partial G_{2} / \partial t }$ and ${ \partial G_{3} / \partial t }$.
These calculations are cumbersome,
and are performed in~\cite{MMA}.
We do not reproduce here these generic equations
that can also be found in Appendix~{B} of~\cite{fbcN2}.

\subsection{Truncating the evolution equations}
\label{sec:Truncation}

To continue the calculation, we may now truncate
the three evolution equations at order ${1/N^{2}}$.
At this stage, the main point is to note that the evolution equation
for ${ \partial F / \partial t }$ only involves ${ G_{2} }$,
whose norm scales like ${ 1/N }$.
As a consequence, in order to derive an equation at order ${ 1/N^{2} }$,
one has to account for the corrections at order ${1/N^{2}}$ that arise in $G_{2}$.
Introducing explicitly the small parameter ${ \epsilon = 1/N }$,
we therefore write
\begin{equation}
G_{2} = \epsilon \, G_{2}^{(1)} + \epsilon^{2} \, G_{2}^{(2)} .
\label{def_G2_expansions}
\end{equation}
Similarly, recalling the definition ${ \mu = \Mtot /N }$,
we can finally perform in the BBGKY equations the replacements
\begin{equation}
\mu \to \epsilon \, \mu , \;\;\; G_{3} \to \epsilon^{2} \, G_{3} , \;\;\; G_{4} \to \epsilon^{3} \, G_{4} .
\label{replacement_scalings}
\end{equation}

At this stage, we are now in a position to truncate the three first BBGKY equations
by keeping only terms up to order $\epsilon^{2}$.
Moreover, relying on the split from Eq.~\eqref{def_G2_expansions},
we also split the evolution equation for ${ \partial G_{2} / \partial t }$
to obtain one evolution equation for ${ \partial G_{2}^{(1)} / \partial t }$ 
(of order ${1/N}$)
and one for ${ \partial G_{2}^{(2)} / \partial t }$
(of order ${ 1/N^{2} }$).

We can further simplify the evolution equations,
by relying on our homogeneous assumptions,
i.e.\@, one has ${ F = F(v , t) }$, independent of $\theta$.
As a result, any term involving ${ \partial F / \partial \theta }$ vanishes.
Similarly, the mean field potential,
${ \!\int\! \rd \bw_{2} \, F(\bw_{2}) U'(\theta_{1} - \theta_{2}) }$,
also vanishes.

In order to ease the analytical derivation of the kinetic equation,
we assume that the system is dynamically hot,
so that the contributions from collective effects can be neglected.
This assumption neglects any backreaction
of a correlation onto the instantaneous potential within which it evolves.
In a nutshell, it neglects integral terms of the form
\begin{equation}
\!\! \int \!\! \rd \bw_{3} \, G_{2}^{(1)} (\bw_{2} , \bw_{3}) \, U' (\theta_{1} - \theta_{3}) \to 0 ,
\label{neglect_collective_effects}
\end{equation}
and similar terms for $G_{2}^{(2)}$ and $G_{3}$.

The last truncations and simplifications that we perform are as follows.
First, in the evolution equation for ${ \partial F / \partial t }$,
we may neglect the source term in $G_{2}^{(1)}$
responsible for the usual ${1/N}$ Landau term,
as it vanishes for 1D homogeneous systems.
Second, in the evolution equation for ${ \partial G_{2}^{(2)} / \partial t }$,
we can neglect the source term in ${ G_{2}^{(1)} }$,
as it does not contribute to the kinetic equation
(see~\cite{MMA}).
Finally, in the evolution equation for ${ \partial G_{3} / \partial t }$,
we can neglect, in the hot limit, the source term in ${ G_{2}^{(1)} \!\times\! G_{2}^{(1)} }$
as its contribution is a factor ${1/Q}$ smaller than the source term
in $G_{2}^{(2)}$.

All in all, as a result of these truncations,
one obtains a set of four coupled differential equations
that describe self-consistently the system's dynamics at order ${1/N^{2}}$.
We do not repeat here these equations which can be found
in Appendix~{C} of~\cite{fbcN2}.

\subsection{Solving the equations}
\label{sec:Solving}

The key property of the previous coupled evolution equations
is that they form a closed and well-posed hierarchy of coupled partial
differential equations.
In particular, because we have neglected collective effects,
there is no need to invert integral operators,
so that the equations can be solved sequentially.
As such, we first solve for the time evolution for ${G_{2}^{(1)}}$,
then $G_{3}$, $G_{2}^{(2)}$, and finally $F$.
At each stage of this calculation, the previous solution
is used as a time-dependent source term in the next evolution equation.

In practice, to solve these equations we rely on Bogoliubov's ansatz,
i.e.\@, we assume ${ F (v , t) = \mathrm{cst.} }$
on the (dynamical) timescale over which correlations evolve.
We also neglect transients associated with initial conditions,
i.e.\@, we solve the evolution equations with the initial conditions
${ G_{2}^{(1)} (t \!=\! 0) \!=\! 0 }$,
and similarly for $G_{2}^{(2)}$ and $G_{3}$.
Finally, in order to describe the process of phase mixing,
we rely on the ${2\pi}$-periodicity of the angle coordinate,
and Fourier expand any function depending on $\theta$,
e.g.\@, following Eq.~\eqref{Fourier_Uk} for the interaction potential.

Having obtained an explicit expression for the time-dependence of ${ G_{2}^{(2)} (t) }$,
we can now aim for the expression of the collision operator ${ \partial F / \partial t }$.
Relying once again on Bogoliubov's ansatz,
this amounts to taking the limit ${ t \!\to\! + \infty }$ in ${ G_{2}^{(2)} (t) }$.
A typical time integral takes the form ${ \!\int_{0}^{t} \! \rd t_{1} \re^{- \ri (t - t_{1}) \omega}}$,
where the frequency $\omega$ is a linear combination of velocities.
Because we have solved three evolution equations sequentially,
we can get up to three such integrals nested in one another,
with partial derivatives w.r.t.\ velocities intertwined in them.
To obtain the asymptotic time behaviour,
we rely on the formula
(see, e.g.\@, Eq.~{(D2)} of~\citep{fbcN2})
\begin{equation}
\lim\limits_{t \to + \infty} \!\! \int_{0}^{t} \!\! \rd t_{1} \, \re^{- \ri (t - t_{1}) \omega} = \pi \deltaD (\omega) - \ri \mP \bigg( \frac{1}{\omega} \bigg) ,
\label{asymp_time}
\end{equation}
with ${ \deltaD (\omega) }$ the Dirac delta,
and ${ \mP (1/\omega) }$ the Cauchy principal value.
It is only at this stage that we evaluate the intertwined gradients
w.r.t.\ the velocities so that they only act on the Dirac deltas
and the Cauchy principal values.

Following all these manipulations, we still have
a kinetic equation involving hundreds of terms,
and requiring further simplifications.
This is the stage where the symbolic algebra system
allows for an efficient manipulation of the formal expressions.
The key steps of these manipulations are:
(i) perform integrations by parts,
so that all the $\deltaD^{\prime}$ and $\deltaD^{\prime\prime}$
are transformed into $\deltaD$;
(ii) use the scaling relations of $\deltaD$ and $\mP$
(and their derivatives),
e.g.\@, ${ \deltaD(\alpha \, x) = \deltaD(x)/|\alpha| }$,
to take out the Fourier wavenumbers as much as possible;
(iii) perform appropriate relabellings of the dummy velocities
and dummy wavenumbers,
so that the sole resonance condition present is
${ \deltaD (k_{1} (v-v_{1}) + k_{2} (v - v_{2})) }$,
i.e.\@, the same resonance condition as in Eq.~\eqref{kin_eq};
(iv) use the presence of the resonance condition,
${ \deltaD (k_{1} (v-v_{1}) + k_{2} (v - v_{2})) }$,
to make the replacements
${ (v\!-\!v_{2}) \to - (k_{1})/(k_{2}) (v-v_{1}) }$
and
${ (v_{1}\!-\!v_{2}) \to - (k_{1}\!+\!k_{2})/(k_{2}) (v-v_{1})}$,
so that the principal values are only expressed as functions of ${(v-v_{1})}$.

After all these cumbersome manipulations,
which we automated using some custom
grammar in \texttt{Mathematica},
one finally obtains the closed result from Eq.~\eqref{kin_eq}.
All the details and functions used for these calculations
can be found in~\cite{MMA}.

\section{Linear Response Theory}
\label{sec:LinearTheory}

When deriving the kinetic equation~\eqref{kin_eq},
we had to neglect the contributions associated
with collective effects.
As a result, this equation only applies in dynamical hot systems,
where the self-consistent amplification of collective effects
is unimportant.
Luckily the amplitude of this dressing of the perturbations
is straightforward to estimate by solving the linear response
theory of the system.

A systematic approach for that calculation is to rely
on already well-established results regarding
the linear stability of inhomogeneous long-range interacting systems.
As detailed in Eq.~{(5.94)} of~\cite{BinneyTremaine2008},
a system's stability is generically governed by the response matrix
\begin{equation}
\wbM_{pq} (\omega) \!=\! 2 \pi \sum_{k} \!\! \int \!\! \rd J \, \frac{k \, \partial F / \partial J}{\omega - k \Omega(J)} \, \psi^{(p)*}_{k} (J) \, \psi^{(q)}_{k} (J) ,
\label{Fourier_M}
\end{equation}
with ${ (\theta , J) = (\theta , v) }$ the angle-action coordinates,
and ${ \Omega (J) = v }$ the orbital frequencies.
In that expression, following the so-called matrix method~\citep{Kalnajs1976},
we introduced a biorthogonal set of basis elements on which the pairwise interaction
is decomposed.
For the present system, the natural basis elements follow from the
Fourier decomposition of the interaction,
that can be written under the separable form
\begin{align}
U (\theta_{1} - \theta_{2}) & \, = - \sum_{p} \psi^{(p)} (\theta_{1}) \, \psi^{(p) *} (\theta_{2}) ,
\nonumber
\\
\psi^{(p)} (\theta) & \, = \sqrt{U_{p}} \, \re^{\ri p \theta} .
\label{def_basis}
\end{align}
The Fourier transform of the basis elements is straightforward to compute.
It is independent of the action $v$,
and reads ${ \psi^{(p)}_{k} = \delta_{p}^{k} \sqrt{U_{k}} }$.
We may finally introduce the dielectric function,
${ \bveps = \bI - \wbM }$,
that is the matrix
\begin{equation}
\veps_{pq} (\omega) = \delta_{p}^{q} \bigg\{ 1 - 2 \pi U_{p} \!\! \int \!\! \rd v \, \frac{p \, \partial F / \partial v}{\omega - p v} \bigg\} .
\label{def_veps}
\end{equation}
As expected for homogeneous systems,
we recover that the dielectric matrix is diagonal,
${ \veps_{pq} (\omega) = \delta_{p}^{q} \, \veps_{p} (\omega) }$,
so that Fourier harmonics are independent from one another.

Using the same dedimensionalisation as in Eq.~\eqref{kin_eq_ddim},
we can rewrite the dielectric function as
\begin{equation}
\veps_{k} (\oom) = 1 - \frac{\oU_{k}}{Q} \!\! \int \!\! \rd u \, \frac{k \, \partial \oF / \partial u}{\oom - k u} ,
\label{veps_ddim}
\end{equation}
with ${ \oom = \omega \td }$ a dimensionless frequency.

In the particular case where the system's \DF\ is single-humped,
i.e.\@, possesses a single maximum,
and is also even, i.e.\@, ${ F (-v) = F(v) }$, so that the maximum
is reached in ${ v = 0 }$,
one can even better characterise the system's dielectric matrix.
In that case, the \DF\ is linearly stable if, and only if,
${ \veps_{k} (0) > 0 }$ for all $k$ (see, e.g.\@,~\cite{fbcN2}).
Following Eq.~\eqref{veps_ddim},
and recalling that the rescaled coupling coefficient
${ \oU_{k} = U_{k} / \Umax }$ is such that ${ | \oU_{k} | \leq 1 }$,
the \DF\ is linearly stable if, and only if, one has
\begin{equation}
Q > \Qc = - \!\! \int \!\! \rd u \, \frac{\partial \oF / \partial u}{u} .
\label{def_Qc}
\end{equation}
One can easily compute the stability limit $\Qc$ for simple \PDFs\@.
In particular, for a Gaussian \PDF\@, one finds ${ \Qc = 1 }$.

\section{Numerical simulations}
\label{sec:NumericalSimulations}

Let us briefly detail the setup of our numerical simulations
used to investigate the long-term relaxation of the Ring model.
Following the Hamiltonian from Eq.~\eqref{def_H_ring},
the equations of motion for particle $i$ reads
\begin{align}
\frac{\rd \theta_{i}}{\rd t} & \, = v_{i} , 
\nonumber
\\
\frac{\rd v_{i}}{\rd t} & \, = - \sum_{j = 1}^{N} \frac{\mu}{2} \frac{\sin (\theta_{i} - \theta_{j})}{(1 - \cos (\theta_{i} - \theta_{j}) + \epsilon)^{3/2}} .
\label{eq_motion_Ring}
\end{align}
We note that in the expression of the acceleration, ${ \rd v_{i} / \rd t }$,
the sum runs over all particles including $i$.
Including this self-interaction is fine here, because the interaction potential
does not diverge at zero separation owing to the softening length, $\epsilon$.
Proceeding in that fashion simplifies the numerical implementation.

Since the Hamiltonian from Eq.~\eqref{def_H_ring} is separable,
one can easily devise symplectic integration schemes for that problem.
In practice, we used the fourth-order symplectic integrator from~\cite{Yoshida1990},
that requires only three (costly) force evaluations per timestep.
However, we note that without any harmonic expansion of the interaction potential,
the equations of motion from Eq.~\eqref{eq_motion_Ring} truly form a $N$-body system,
as the computation of each acceleration requires ${ \mO (N) }$ operations.

In order to accelerate the integration of that system,
we followed an approach similar to~\cite{RochaFilho2014},
and implemented the computations on \GPUs\@.
In practice, simulations were run on \texttt{NVIDIA V100} \GPUs\@,
with ${ N = 1024 }$ particles per simulation,
and ${ \Nthreads = N }$ threads per computation block,
i.e.\@, one thread per particle.
For this particular \GPU\@,
we could run ${ \Nblocks = 80 }$ independent realisations
simultaneously on a given \GPU\@.
In total, we performed ${ \Nruns = 20 }$ different batches
of simulations,
i.e.\@, we had a total of ${1600}$ independent realisations
to perform the ensemble average.

In the numerical implementation, the computation of the particles' accelerations
is by far the most numerically demanding task.
To accelerate these evaluations, we focused on three main points.
(i) First, in Eq.~\eqref{eq_motion_Ring}, the trigonometric functions
${ \cos (\theta_{i} \!-\! \theta_{j}) }$ and ${ \sin (\theta_{i} - \theta_{j}) }$
are expanded using duplications formulae,
so that one only has to compute ${ (\sin(\theta_{i}) , \cos (\theta_{i})) }$
for every particle, using the instruction \texttt{sincos}.
(ii) Second, these harmonic functions are pre-computed once per particle,
and loaded in shared data array to allow for fast coalesced memory accesses
for all the threads in the same computation block.
(iii) Third, the computation of the force in Eq.~\eqref{eq_motion_Ring}
was further accelerated by using the instruction \texttt{rsqrt(x)}
that allows for a fast computation of ${ 1/\sqrt{x} }$.
With such parameters, integrating for one timestep required ${1.3 \,\mathrm{ms}}$
of computation time.

In practice, we set the softening length to ${ \epsilon = 0.01 }$,
which imposes ${ \Umax \simeq 0.92 }$,
as defined in Eq.~\eqref{def_Q}.
We used an integration time step equal to ${ \delta t = 1/(50 \!\times\! \sigma) }$,
that guaranteed a relative error in the total energy
of the order of ${10^{-5}}$.
Each realisation was integrated for a total of ${ 4\!\times\!10^{8} }$
timesteps, requiring about 6 days of computation per realisation.

We used the same initial conditions as in~\cite{fbcN2},
given by a generalised Gaussian distribution following
\begin{align}
P (v) & \, = \frac{\alpha}{2} \, \frac{A (\alpha , \sigma)}{\Gamma (1 / \alpha)} \, \exp \big[ - (A(\alpha , \sigma) |v|)^{\alpha} \big] ,
\nonumber
\\
A (\alpha , \sigma) & \, = \frac{1}{\sigma} \, \bigg( \frac{\Gamma (3/\alpha)}{\Gamma (1/\alpha)} \bigg)^{1/2} .
\label{def_nonGaussian}
\end{align}
This \PDF\ is normalised so that ${ \!\int\! \rd v P (v) = 1 }$,
is of zero mean, and of variance $\sigma^{2}$.
The particular case ${ \alpha = 2 }$ corresponds to the case of the Gaussian distribution,
already introduced in Eq.~\eqref{def_Boltzmann},
whose stability threshold,
following Eq.~\eqref{def_Qc},
reads ${ \Qc = 1 }$.
In practice, in the numerical simulations,
we used the value ${ \alpha = 4 }$,
which corresponds to a less peaked \PDF\@,
and chose the initial velocity dispersion to be ${ \sigma = 3 }$.
Finally, assuming ${ \Mtot = 1 }$, the stability parameter, $Q$,
from Eq.~\eqref{def_Q} becomes ${ Q \simeq 9.75 }$,
while the stability threshold is ${ \Qc \simeq 0.46 }$
(see~\citep{fbcN2}),
i.e.\@, the considered initial condition is linearly stable.
To measure in Fig.~\ref{fig:Flux}
the diffusion flux and the associated errors
(16\% and 84\% confidence levels),
we followed the exact same procedure as detailed
in Appendix~{F} of~\cite{fbcN2}.
We do not repeat it here.

\begin{figure}[htbp!]
\begin{center}
\includegraphics[width=0.48\textwidth]{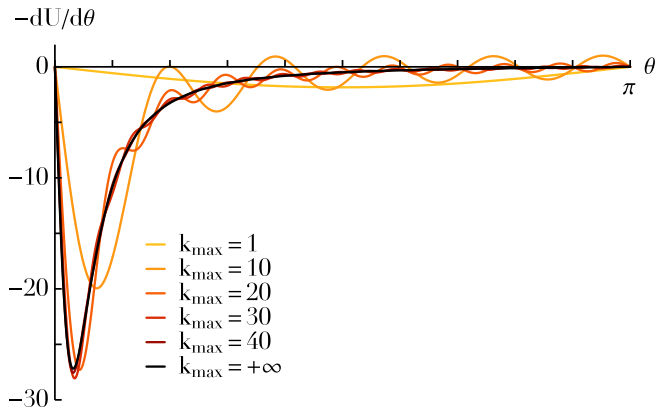}
\caption{Illustration of the pairwise force, ${ - \rd U / \rd \theta }$,
as one varies the maximum index $\kmax$ considered in the
interaction potential.
As expected, the larger $\kmax$, the better the reconstruction
of the exact interaction.
\label{fig:dUdthetakmax}}
\end{center}
\end{figure}

\begin{figure}[hbtp!]
\begin{center}
\includegraphics[width=0.48\textwidth]{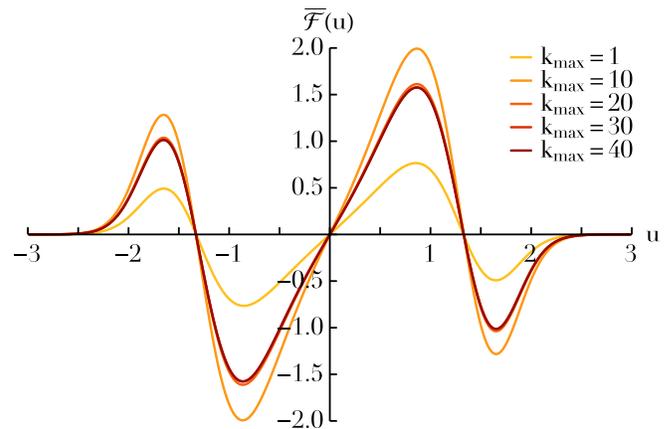}
\caption{Illustration of the dependence of the dimensionless flux,
${ \omF (u) }$,
as one varies the maximum index ${\kmax}$ considered
in the interaction potential.
The considered system and initial conditions
are identical to Fig.~\ref{fig:Flux}.
As soon as the truncated Fourier series of $U_{k}$
represents accurately enough the underlying potential, ${ U (\theta_{1} \!-\! \theta_{2}) }$,
the kinetic predictions have converged.
\label{fig:Predkmax}}
\end{center}
\end{figure}

The kinetic equation~\eqref{kin_eq}
involves an infinite sum over $k_{1},k_{2}$.
In pratice, one has to truncate these sums.
To do so,
we may truncate the interaction potential from Eq.~\eqref{Fourier_Uk},
so that ${ U_{k} = 0 }$ for ${ |k| > \kmax }$.
The effect of such a truncation
on the pairwise force is illustrated in Fig.~\ref{fig:dUdthetakmax}.
Doing so, we may then restrict the sums over fundamental resonances,
as defined in Eq.~\eqref{rewrite_kin_eq},
only to ${ 0 < k , \kp \leq \kmax }$.
Figure~\ref{fig:Predkmax}
illustrates the effect of $\kmax$ on the computed diffusion flux.
In that figure, we recover that for $\kmax$ large enough,
the diffusion flux converges,
so that higher order resonances do not contribute anymore to the relaxation.
In practice, for the considered softening ${ \epsilon = 0.01 }$,
we used ${ \kmax = 40 }$ in the predictions from Fig.~\ref{fig:Flux}.

\end{document}